\documentclass[preprint]{elsarticle} 
\usepackage{graphicx}
\begin{document}
\title{Self consistent charge-current in a mesoscopic region attached to
       superconductor leads}

\author{D.\ Verrilli} 
\ead{dverrill@fisica.ciens.ucv.ve}

\author{F.\ P.\ Marin} 
\ead{felix@fisica.ciens.ucv.ve}
\address{Laboratorio de F\'{\i}sica Te\'{o}rica de S\'{o}lidos (LFTS).
         Centro de F\'{\i}sica Te\'{o}rica y Computacional (CEFITEC).
         Facultad de Ciencias. Universidad Central de Venezuela.
         A.P. 47586. Caracas 1041-A. Venezuela}

\begin{abstract}
We investigate the behavior of an electric potential profile inside a
mesoscopic region attached to a pair of superconducting leads. It turns
out that the ${\rm I}-V$ characteristic curves are strongly modified by
this profile. In addition, the electronic population in the mesoscopic
region is affected by the profile behavior. We derive the single
particle current and the mesoscopic electronic population within the
non-equilibrium Keldysh Green functions technique. The Keldysh technique
results are further converted in a self consistent field (SCF) problem
by introducing potential profile modifications. Evaluation of ${\rm I}-V$ 
characteristics are presented for several values of the
model parameters and comparison with current experimental results are
discussed.
\end{abstract}

\begin{keyword}
Quantum dot, Superconductors, Keldysh Green Functions
\end{keyword}

\maketitle

\section{Introduction}\label{intro}
From the demostration of a superconductor-normal-super\-con\-duc\-tor
(S-N-S) transistor \cite{Baselmans}, the study of the
non\-equi\-li\-brium transport through superconducting systems has been
of much interest \cite{Volkov,Wilhelm,Yip,Giazotto,Pala}. Another
interesting problem in mesoscopic physics is transport through a
su\-per\-con\-duc\-tor\-/quan\-tum dots/superconductor system
\cite{Defranceschi,Fazio,Fazio1,PhysRevB.57.11891,Schwab,Clerk,Shapira,
PhysRevB.55.R6137,Cuevas}. In this paper, we study the effect of an
electrostatic potential profile on the electric transport across a
single quantum dot with a spin degenerated level. Such a dot is coupled
to a pair of biased superconductors contacts or leads (source and
drain). By applying a source voltage $V_{S}$ and a drain voltage $V_{D}$
an electric current can flow between the leads and across the quantum
dot which sets a typical non equilibrium situation. Besides the applied
drain voltages $V_{D}$ and source voltage $V_{S}$ the system is further
manipulated by a gate voltage $V_{G}$ which, in principle, couples
directly to the quantum dot. It turns out that $V_{D}$, $V_{S}$ and
$V_{G}$ induce an effective electrostatic profile potential inside the
mesoscopic region in such a way that electronic population and electric
current become tied to a self consistent problem. It is quite clear that
such potential profile modifies the quantum dot level structure in a
self consistent fashion. Such situation can be highly complicated since
it mixes non equilibrium statistical mechanics with a classic
electrostatic framework. Here, we adopt an approach which relates the
self consistent electrostatic profile to the electronic population of
the quantum dot and to the electric current\cite{Liang,datta95}. The
self consistency and any other model calculations are fully performed
within the non equilibrium Keldysh technique \cite{keldysh65,haug}.
 
In section \ref{calc} we find the expression for the current and the
electronic population for a mesoscopic region. In addition, we show
calculations which lead to a self consistent field (SCF) problem between
the dot electronic population and the electric current between the
superconducting leads. The self consistency takes into account electric
potential profiles inside the mesoscopic region as induced by the drain
and source bias and by the gate voltage \cite{datta95}. Moreover we
present the results about the effect of the potential profile on the
${\rm I}-V$ characteristic curves and on the electronic population
inside the mesoscopic region.

Finally, in section \ref{conclusions} we discuss our main conclusions.
%%%%%%%%%%%%%%%%%%%%%
\section{Calculation}
\label{calc}
In this section we present the model and calculations which lead to the
current and to the population number in the mesoscopic region.

We consider a spin degenerated single orbital in a quantum dot connected
to superconductors leads. The hamiltonian which describes this system is
a generalized Friedel-Anderson model \cite{PhysRev.124.41}. It reads
\begin{equation}
H = H_{S} + H_{D} + H_{T},
\label{equation1}
\end{equation}
where $H_{S}$, $H_{D}$ and $H_{T}$ stand for the superconducting leads,
the dot and the tunneling term, respectively. $H_{S} = H_{L} + H_{R}$
where $H_{L}$ and $H_{R}$ are the left and right lead hamiltonians,
respectively. They are given, within the BCS model
\cite{PhysRev.108.1175}, by
\begin{equation}
H_{S}
=
\sum_{\eta\vec{k}\sigma}
\Psi^{\dagger}_{\eta\vec{k}\sigma}
\mathrm{H}^{0}_{\eta\vec{k}}
\Psi_{\eta\vec{k}\sigma}
\label{equation2}
\end{equation}
with
\begin{equation}
\mathrm{H}^{0}_{\eta\vec{k}}
=
\left(%
\begin{array}{cc}
\varepsilon_{\eta\vec{k}}&\Delta_{\eta\vec{k}}
\\
\Delta^{*}_{\eta\vec{k}}&-\varepsilon_{\eta\vec{k}}
\end{array}\right)
\label{equation3}
\end{equation}
where $\Delta_{\eta\vec{k}}$ is the superconductor gap,  of the lead
$\eta = L, R$. $\Psi^{\dagger}_{\eta\vec{k}\sigma}$ and
$\Psi_{\eta\vec{k}\sigma}$ are the Nambu spinors
\begin{equation}
\Psi_{\eta\vec{k}\sigma}^{\dagger}
=
\left(%
a_{\eta\vec{k}\sigma}^{\dagger}
\quad
a_{\eta,-\vec{k},-\sigma}\right),
\qquad
\Psi_{\eta\vec{k}\sigma}
=
\left(%
\begin{array}{c}
a_{\eta\vec{k}\sigma}
\\
a^{\dagger}_{\eta,-\vec{k},-\sigma}
\end{array}\right)
\label{equation4}
\end{equation}
$H_{D}$ is the hamiltonian for the single-level quantum
dot  of energy $E_{d}$:
\begin{equation}
H_{D}
= 
\sum_{\sigma}\phi^{\dagger}_{\sigma}\mathrm{H}^{QD}\phi_{\sigma}.
\label{equation5}
\end{equation}
with
\begin{equation}
\mathrm{H}^{QD}
=
\left(%
\begin{array}{cc}
\mathrm{E}_{d} + \mathrm{U}_{d}n_{-\sigma} & 0
\\
0 & -\mathrm{E}_{d} - \mathrm{U}_{d}n_{\sigma}
\end{array}\right)
\label{equation6}
\end{equation}
The tunneling hamiltonian $H_{T}$ is given by
\begin{equation}
H_{T}
=
\sum_{\eta\vec{k}\sigma}
\Psi_{\eta\vec{k}\sigma}^{\dagger}
\mathrm{H}^{I}_{\eta\vec{k}}\phi_{\sigma}
\label{equation7}
\end{equation}
with
\begin{equation}
\mathrm{H}^{I}_{\eta\vec{k}}
=
\left(%
\begin{array}{cc}
V_{\eta\vec{k}} & 0
\\
0 & -V_{\eta\vec{k}}
\end{array}\right)
\label{equation8}
\end{equation}
$H_{T}$ connects the dot to the biased superconducting leads and it
allows the electric charge flow. $V_{\eta\vec{k}}$ is the hybridization
matrix element between a conduction electron of energy
$\varepsilon_{\eta\vec{k}}$ in the $\eta = L, R$ superconductor lead and
a localized electron on the dot with energy $E_{d}$.
$\phi_{\sigma}^{\dagger}$ and $\phi_{\sigma}$ are the dot spinors
\begin{equation}
\phi_{\sigma}^{\dagger}
=
\left(d_{\sigma}^{\dagger}\quad d_{-\sigma}\right),
\qquad
\phi_{\sigma}
=
\left(%
\begin{array}{c}
d_{\sigma}
\\
d^{\dagger}_{-\sigma}
\end{array}\right)
\label{equation9}
\end{equation}
Here
$a^{\dagger}_{\eta\vec{k}\sigma}\left(a_{\eta\vec{k}\sigma}\right)$
denotes the creation (annihilation) operator for a conduction electron
with the wave vector $\vec{k}$, spin $\sigma$ in the $\eta = L, R$
superconductor lead. $d^{\dagger}_{\sigma}\left(d_{\sigma}\right)$ is
the creation (annihilation) operator for an electron on the dot.

The flow of electric charge from the terminal $\eta$ is given by
\begin{equation}
{\rm I}_{\eta}\left( t\right)
=
\left(-e\right)
\left\lbrack
-\frac{d\left\langle N_{\eta}\left(t\right)\right\rangle}{dt}
\right\rbrack
=
\frac{{\rm i} e}{\hbar}
\left\langle
\left\lbrack H_{T}\left(t\right), N_{\eta}\left(t\right)\right\rbrack
\right\rangle,
\label{equation10}
\end{equation}
where $-{\rm e}$ is the electron charge.
$\left\langle\cdots\right\rangle$ is the thermodynamical average over
the biased $L$ and $R$ leads at the temperature $T$. Equation
(\ref{equation10}) can be expressed in terms of the Keldysh Green
function
\begin{equation}
{\rm F}_{\eta\vec{k}\sigma}\left(t, t'\right)
\equiv 
-{\rm i}\langle%
{\rm T_{c}}a_{\eta\vec{k}\sigma}\left(t\right)
           d^{\dagger}_{\sigma}\left(t'\right)\rangle
\label{equation11}
\end{equation}
as
\begin{equation}
{\rm I}_{\eta}\left(t\right)
=
\frac{2e}{\hbar}V_{\eta}\Re\sum_{\vec{k}\sigma}
{\rm F}^{<}_{\eta\vec{k}\sigma}\left(t, t\right),
\label{equation12}
\end{equation}
where ${\rm F}^{<}_{\eta\vec{k}\sigma}\left(t, t'\right)$ is a lesser
Keldysh Green function. For the purpose of the single particle current
evaluation the coupling~$\left\vert V_{\eta\vec{k}}\right\vert^{2}$ can
be replaced by an average~$V_{\eta}^{2}$ at the Fermi surfaces of the
leads $L$ and $R$. Hereafter, for simplicity, we replace
$V_{\eta\vec{k}}$ by $V_{\eta}$ as we already do it in eqn
(\ref{equation12}).

The first evaluation step of
${\rm F}^{<}_{\eta\vec{k}\sigma}\left(t, t'\right)$ expresses it in
terms of dot Keldysh Green functions. Then, we set an equation
of motion for the Keldysh Green function
${\rm F}_{\eta\vec{k}\sigma}\left(t, t'\right)$
\begin{equation}
\left({\rm i}\frac{\partial}{\partial t}
      -
      \epsilon_{\eta\vec{k}}\right)
{\rm F}_{\eta\vec{k}\sigma}\left(t, t'\right)
=
-\sigma\Delta_{\eta}{\mathcal F}_{\eta\vec{k}\sigma}\left(t, t'\right)
+
V_{\eta}{\rm G}_{\sigma}\left(t, t'\right),
\label{equation13}
\end{equation}
where
\begin{eqnarray}
{\mathcal F}_{\eta\vec{k}\sigma}\left(t, t'\right)
& = &
-{\rm i}
\left\langle
{\rm T_{c}}
a^{\dagger}_{\eta\vec{k},-\sigma}\left(t\right)
d^{\dagger}_{\sigma}\left(t'\right)
\right\rangle, 
\label{equation14}
\\
{\rm G}_{\sigma}\left(t, t'\right)
& = &
-{\rm i}
\left\langle
{\rm T_{c}}d_{\sigma}\left(t\right)
           d^{\dagger}_{\sigma}\left(t'\right)
\right\rangle.
\label{equation15}
\end{eqnarray}

Similarly, ${\mathcal F}_{\eta\vec{k}\sigma}\left(t, t'\right)$
satisfies the equation of motion
\begin{equation}
\left(%
{\rm i}\frac{\partial}{\partial t}
       +
       \epsilon_{\eta\vec{k}}\right)
{\mathcal F}_{\eta\vec{k}\sigma}\left(t, t'\right)
=
-\sigma\Delta_{\eta}{\rm F}_{\eta\vec{k}\sigma}\left(t, t'\right)
-
V_{\eta}{\mathcal G}_{\sigma}\left(t, t'\right),
\label{equation16}
\end{equation}
where
\begin{equation}
{\mathcal G}_{\sigma}\left(t, t'\right)
=
-{\rm i}\left\langle
{\rm T_{c}}d^{\dagger}_{-\sigma}\left(t\right)
           d^{\dagger}_{\sigma}\left(t'\right)
\right\rangle.
\label{equation17}
\end{equation}
The eqns (\ref{equation13}) and (\ref{equation16}) can be written
as follows:
\begin{equation}
\left(%
\begin{array}{cc}
{\rm i}\frac{\partial}{\partial t} - \epsilon_{\eta\vec{k}}
&
\sigma\Delta_{\eta}
\\
\sigma\Delta_{\eta}
&
{\rm i}\frac{\partial}{\partial t} + \epsilon_{\eta\vec{k}}
\end{array}\right)
\left(%
\begin{array}{c} 
{\rm F}_{\eta\vec{k}\sigma}\left(t, t'\right)
\\
{\mathcal F}_{\eta\vec{k}\sigma}\left(t, t'\right)
\end{array}\right)
=
V_{\eta}\sigma_{z}\left(%
\begin{array}{c} 
{\rm G}_{\sigma}\left(t, t'\right)
\\
{\mathcal G}_{\sigma}\left(t, t'\right) \end{array}\right).
\label{equation18}
\end{equation}
This equation can be written as an integral along the Keldysh contour
${\rm C_{K}}$
\begin{eqnarray}
\left(%
\begin{array}{c} 
{\rm F}_{\eta\vec{k}\sigma}(t,t')
\\
{\mathcal F}_{\eta\vec{k}\sigma}(t,t') 
\end{array}\right)
& = &
\int_{\rm C_{K}}{\rm d}t''
\left(%
\begin{array}{cc}
{\rm g}_{\eta\vec{k}\sigma}\left(t, t''\right)
&
\widetilde{\rm f}_{\eta\vec{k}\sigma}\left(t, t''\right)
\\
{\rm f}_{\eta\vec{k}\sigma}\left(t, t''\right)
&
\widetilde{\rm {g}}_{\eta\vec{k}\sigma}\left(t, t''\right) 
\end{array}\right)\times
\nonumber\\
&&
V_{\eta}\sigma_{z}
\left(%
\begin{array}{c} 
{\rm G}_{\sigma}\left(t'', t'\right)
\\
{\mathcal G}_{\sigma}\left(t'', t'\right) 
\end{array}\right),
\label{equation19}
\end{eqnarray}
The $2\times 2$ matrix in the right hand side of eqn
(\ref{equation19}) is an unperturbed Keldysh Green function where
\begin{eqnarray}
{\rm g}_{\eta\vec{k}\sigma}\left(t, t'\right) 
& \equiv &
-{\rm i}
\left\langle
{\rm T_{c}}a_{\eta\vec{k}\sigma}\left(t\right)
           a^{\dagger}_{\eta\vec{k}\sigma}\left(t'\right)
\right\rangle_{0}, 
\label{equation20}
\\
\widetilde{\rm {g}}_{\eta\vec{k}\sigma}\left(t, t'\right)
& \equiv &
-{\rm i}
\left\langle
{\rm T_{c}}a^{\dagger}_{\eta,-\vec{k},-\sigma}\left(t\right)
           a_{\eta,-\vec{k},-\sigma}\left(t'\right)
\right\rangle_{0},
\label{equation21}
\\
\widetilde{\rm f}_{\eta\vec{k}\sigma}\left(t, t'\right)
& \equiv &
-{\rm i}
\left\langle
{\rm T_{c}}a_{\eta\vec{k}\sigma}\left(t\right)
           a_{\eta,-\vec{k},-\sigma}\left(t'\right)
\right\rangle_{0},
\label{equation22}
\\
{\rm f}_{\eta\vec{k}\sigma}\left(t, t'\right)
&\equiv&
-{\rm i}
\left\langle
{\rm T_{c}}a^{\dagger}_{\eta,-\vec{k},-\sigma}\left(t\right)
           a^{\dagger}_{\eta\vec{k}\sigma}\left(t'\right)
\right\rangle_{0}.
\label{equation23}
\end{eqnarray}
The subindex $_{0}$ indicates that evaluations are performed with
$V_{\eta} = 0$.

The contribution ${\rm F}_{\eta\vec{k}\sigma}^{<}\left(t, t'\right)_{SP}$
to the single particle (SP) current is given by

\begin{eqnarray}
{\rm F}_{\eta\vec{k}\sigma}\left(t, t'\right)_{SP}
& = &
V_{\eta}\int_{-\infty}^{\infty}{\rm d}t''\times\nonumber
\\
&&
\left\lbrack%
{\rm g}_{\eta\vec{k}\sigma}^{\rm\left(r\right)}\left(t, t''\right)
{\rm G}_{\sigma}^{<}\left(t'', t'\right)
+
{\rm g}_{\eta\vec{k}\sigma}^{<}\left(t, t''\right)
{\rm G}_{\sigma}^{\rm\left(a\right)}\left(t'', t'\right)
\right\rbrack
\label{equation24}
\end{eqnarray}
where we used eqn (\ref{equation19}) and Langreth rules
\cite{langreth}. The superscripts
$^{<, >, {\rm\left(r\right)}, {\rm\left(a\right)}}$ correspond to
lesser, greater, retarded and advanced  Keldysh Green functions,
respectively. Therefore, the single particle current
${\rm I}_{\eta}\left(t\right)_{SP}$ can be written as
\begin{equation}
{\rm I}_{\eta}\left(t\right)_{SP}
=
\frac{2e}{\hbar}\Re\sum_{\sigma}\int_{-\infty}^{\infty}{\rm d}t'
\left\lbrack%
\Sigma_{\eta\sigma}^{\rm\left(r\right)}\left(t, t'\right)
{\rm G}_{\sigma}^{<}\left(t', t\right)
+
\Sigma_{\eta\sigma}^{<}\left(t, t'\right)
{\rm G}_{\sigma}^{\rm\left(a\right)}\left(t', t\right)
\right\rbrack
\label{equation25}
\end{equation}
$\Sigma_{\eta\sigma}^{\rm\left(r\right)}\left(t, t'\right)
 =
 V_{\eta}^{2}\sum_{\vec{k}}
 {\rm g}_{\eta\vec{k}\sigma}^{\rm\left(r\right)}\left(t, t'\right)$
and
$\Sigma_{\eta\sigma}^{<}\left(t, t'\right)
=
V_{\eta}^{2}\sum_{\vec{k}}
{\rm g}_{\eta\vec{k}\sigma}^{<}\left(t, t'\right)$ are self energies
which are evaluated for isolated superconductors leads $L$ y $R$. They
depend on $t$ and $t'$ through $t - t'$ and are independent of $\sigma$.
Their Fourier transforms are given by
\begin{eqnarray}
\Sigma^{\rm\left(r\right)}_{\eta}(\omega)
& = &
-{\rm \Gamma}_{\eta}\left\lbrack%
\frac{\omega - \mu_{\eta}}{\Delta_{\eta}}
\zeta(\Delta_{\eta},\omega - \mu_{\eta})
+
{\rm i}\zeta(\omega - \mu_{\eta},\Delta_{\eta})\right\rbrack
\label{equation26}
\\
{\rm \Sigma}^{<}_{\eta}(\omega)
& = &
2{\rm i}{\rm \Gamma}_{\eta}
\zeta(\omega - \mu_{\eta},\Delta_{\eta}){\rm f}(\omega - \mu_{\eta})
\label{equation27}
\end{eqnarray}
where
\begin{equation}
\zeta\left(\omega, \omega'\right)
\equiv
\Theta\left(\left\vert\omega\right\vert
            -
            \left\vert\omega'\right\vert\right)
\frac{\left\vert\omega\right\vert}{\sqrt{\omega^{2}-\omega'^{2}}}.
\label{equation28}
\end{equation}
$\Gamma_{\eta} = \pi N_{\eta}\left(0\right)V_{\eta}^{2}$
are the coupling constants between the leads and the
quantum dot in the wide band limit. $N_{\eta}\left(0\right)$ is the
density of states at the $\eta$ Fermi level and
$\mathrm{f}\left(\omega\right)$ is the Fermi-Dirac distribution function.

Equation (\ref{equation25}) becomes
\begin{eqnarray}
{\rm I}_{\eta}\left(t\right)_{SP}
& = &
\frac{2e}{h}\Re
\int_{-\infty}^{\infty}{\rm d}\omega
\int_{-\infty}^{\infty}\frac{{\rm d}\omega'}{2\pi}
\mathrm{e}^{-{\rm i}\left(\omega - \omega'\right)t}\times
\nonumber\\
&&
\left\lbrack%
\Sigma_{\eta}^{\rm\left(r\right)}\left(\omega\right)
\sum_{\sigma}{\rm G}_{\sigma}^{<}\left(\omega, \omega'\right)
+
\Sigma_{\eta}^{<}\left(\omega\right)
\sum_{\sigma}
{\rm G}_{\sigma}^{\rm\left(a\right)}\left(\omega, \omega'\right)
\right\rbrack
\label{equation29}
\end{eqnarray}

Dot Keldysh Green's functions
${\rm G}_{\sigma}^{<}\left(\omega,\omega'\right)$ and
${\rm G}_{\sigma}^{{\rm\left(a\right)}}\left(\omega,\omega'\right)$
are straightforward evaluated. It turns out that they are $\sigma$
independent and frequency diagonal in the stationary limit

\begin{eqnarray}
{\rm G}_{\sigma}^{<}\left(\omega,\omega'\right)
& \equiv &
2\pi\delta\left(\omega - \omega'\right){\rm G}^{<}\left(\omega\right)
\\
{\rm G}_{\sigma}^{{\rm\left(a\right)}}\left(\omega,\omega'\right)
& \equiv &
2\pi\delta\left(\omega - \omega'\right)
{\rm G}^{{\rm\left(a\right)}}\left(\omega\right),
\end{eqnarray}

Equation (\ref{equation29}) becomes
\begin{eqnarray}
{\rm I}_{\eta SP}
& = &
\frac{4e}{h}\Re
\int_{-\infty}^{\infty}{\rm d}\omega
\left\lbrack%
\Sigma_{\eta}^{\rm\left(r\right)}\left(\omega\right)
{\rm G}^{<}\left(\omega\right)
+
\Sigma_{\eta}^{<}\left(\omega\right)
{\rm G}^{\rm\left(a\right)}\left(\omega\right)
\right\rbrack
\label{equationSP0}
\end{eqnarray}

The final expression for the single particle current
${\rm I}_{SP} \equiv \left({\rm I}_{R,SP} - {\rm I}_{L,SP}\right)/2$ is
given by
\begin{equation}
{\rm I}_{SP}
=
\frac{8\pi e}{h}
\int^{\infty}_{-\infty}{\rm d}\omega\,
\frac{\Gamma_{L}\left(\omega - eV\right)\Gamma_{R}\left(\omega\right)}
      {\Gamma_{L}\left(\omega - eV\right) 
       +
      \Gamma_{R}\left(\omega\right)}\,
\rho\left(\omega\right)
\left\lbrack
{\rm f}\left(\omega - eV\right) - {\rm f}\left(\omega\right)
\right\rbrack ,
\label{equation32}
\end{equation}

In eqn (\ref{equation32}) we performed a trivial shift of the dot energy
level and insert the electric potencial $V$ through
$eV = \mu_{L} - \mu_{R}$. The extra $2\pi$ factor arises from the dot
Keldysh Green functions. ${\rm {\rm \Gamma}}_{\eta}\left(\omega\right)$
and $\rho\left(\omega\right)$ are given by
\begin{eqnarray}
\Gamma_{\eta}\left(\omega\right)
& = &
{\rm {\rm \Gamma}}_{\eta}\zeta\left(\omega,\Delta_{\eta}\right)
\\
\rho\left(\omega\right)
& = &
-\,\frac{1}{\pi}
\Im{\rm G}^{\rm\left(r\right)}\left(\omega\right)\nonumber
=
\frac{{\rm {\rm \Gamma}}\left(\omega\right)/\pi}
     {\left(\omega - E_{d}\right)^{2}
      +
      {\rm {\rm \Gamma}}^{2}\left(\omega\right)}
\\
{\rm {\rm \Gamma}}\left(\omega\right)
& = &
{\rm {\rm \Gamma}}_{L}\left(\omega - eV\right)
+
{\rm {\rm \Gamma}}_{R}\left(\omega\right)
\label{equation33}
\end{eqnarray}

Here $\rho\left(\omega\right)$ is the so-called quantum dot spectral
function which is given in terms of the retarded
$\mathrm{G^{\left(r\right)}}\left(\omega\right)$ Keldysh Green function
\cite{keldysh65}. At steady state there is no net flow into or out of
the mesoscopic channel which yields a stationary particle number in it.
The population number $N$, at the dot, is given by

\begin{equation}
N
=
2\left\lbrack -{\rm i}{\rm G}^{<}\left(t,t\right)\right\rbrack
=
2\int^{\infty}_{-\infty}\frac{{\rm d}\omega}{2\pi{\rm i}}\,
{\rm G}^{<}\left(\omega\right),
\label{equation34}
\end{equation}
which becomes a weighted average over the $L$ and $R$ contacts
\begin{equation}
N
=
2\int^{\infty}_{-\infty}{\rm d}\omega\,\rho\left(\omega\right)
\left\lbrack
\frac{\Gamma_{L}\left(\omega - eV\right)}{\Gamma\left(\omega\right)}\,
{\rm f}\left(\omega - eV\right)
+ 
\frac{\Gamma_{R}\left(\omega\right)}{\Gamma\left(\omega\right)}\,
{\rm f}\left(\omega\right)
\right\rbrack.
\label{equation35}
\end{equation}

So far, we are not included the side effects of a potential profile
inside the mesoscopic channel. Such potential is induced by the action
of source, drain and gate applied voltages. Since the number of quantum
levels in the channel is small the particle number variation is
negligible. It amounts to neglect potential profile variation inside the
channel. Then we can visualize the channel as a single point and an
equivalent circuit framework is quite useful. In this framework we
associate capacitances $C_{D}$, $C_{S}$ y $C_{G}$ to the drain, source
and gate, respectively. Whenever drain, source and gate bias potentials
$V_{D}$, $V_{S}$ and $V_{G}$, respectively, are present it induces a
shift $U = -e\left(V_{ch} - V_{0}\right)$ of the electrostatic energy
inside the channel. $V_{ch}$ and $V_{0}$ are channel electrostatic
potentials after and before we apply the source and drain biases,
respectively. The electronic population before and after we apply the
biases mentioned above are given by
\begin{eqnarray}
-eN_{0}
& = &
C_{D}V_{0} + C_{S}V_{0} + C_{G}V_{0}
\\
-eN
& = &
C_{D}\left(V_{ch} - V_{D}\right) + C_{S}\left(V_{ch} - V_{S}\right)
+
C_{G}\left(V_{ch} - V_{G}\right),
\end{eqnarray}   
respectively. It leads us to
\begin{equation}
-e\Delta N
\equiv
-e\left(N - N_{0}\right)
=
C_{E}\left(V_{ch} - V_{0}\right) - C_{D}V_{D} - C_{S}V_{S} - C_{G}V_{G}
\end{equation}
where $C_{E} = C_{D} +  C_{S} + C_{G}$. The electrostatic potencial
shift $U$ inside the channel becomes
\begin{equation}
U = U_{\cal L} + \frac{e^{2}}{C_{E}}\,\Delta N
\label{equation36}
\end{equation}
where
\begin{equation}
C_{E}U_{\cal L}
\equiv
C_{D}\left(-eV_{D}\right)
+
C_{S}\left(-eV_{S}\right)
+
C_{G}\left(-eV_{G}\right)
\label{equation36}
\end{equation}
The first term yields linear contributions to the potencial profile
while the second one introduces a direct dependence on the electronic
population $N$. $U_{0} = e^{2}/C_{E}$ is the dot charging energy.
$C_{E}$ is an effective dot capacitance which depends on drain $C_{D}$,
source $C_{S}$ and gate $C_{G}$ capacitances within an equivalent
circuit framework. 

The potential profile $U$ shifts the dot quantum levels such that
${\rm I}_{SP}$ and $N$ are found from a system of self consistent
equations.
\begin{equation}
{\rm I}_{SP}
=
\frac{8\pi e}{h}
\int^{\infty}_{-\infty}{\rm d}\omega\,
\frac{\Gamma_{L}\left(\omega - eV\right)\Gamma_{R}\left(\omega\right)}
     {\Gamma_{L}\left(\omega - eV\right)
      +
      \Gamma_{R}\left(\omega\right)}\,
\rho\left(\omega - U\right)
\left\lbrack
{\rm f}\left(\omega - eV\right) - {\rm f}\left(\omega\right)
\right\rbrack ,
\label{equation39}
\end{equation}

\begin{equation}
N
=
2\int^{\infty}_{-\infty}{\rm d}\omega\,\rho\left(\omega - U\right)\,
\frac{\Gamma_{L}\left(\omega - eV\right)
      {\rm f}\left(\omega - eV\right)
      +
      \Gamma_{R}\left(\omega\right){\rm f}\left(\omega\right)}
      {\Gamma_{L}\left(\omega - eV\right)
      +
      \Gamma_{R}\left(\omega\right)} .
\label{equation40}
\end{equation}

Equation (\ref{equation40}) determines $N$ in a self consistent fashion
which inmediately yields the single particle electric current
${\rm I}_{SP}$ by carrying out the integration in eqn (\ref{equation39}).

We will consider a situation where the lead couplings are not extremely
small and the dot capacitance is reasonably large. It will smear out the
Coulomb blockade effect and the double occupancy of the resonant level
will be very unlikely.

\section{Results and Discussion}
\label{ResandDisc}
\begin{figure}
\resizebox{1.0\hsize}{!}{\includegraphics{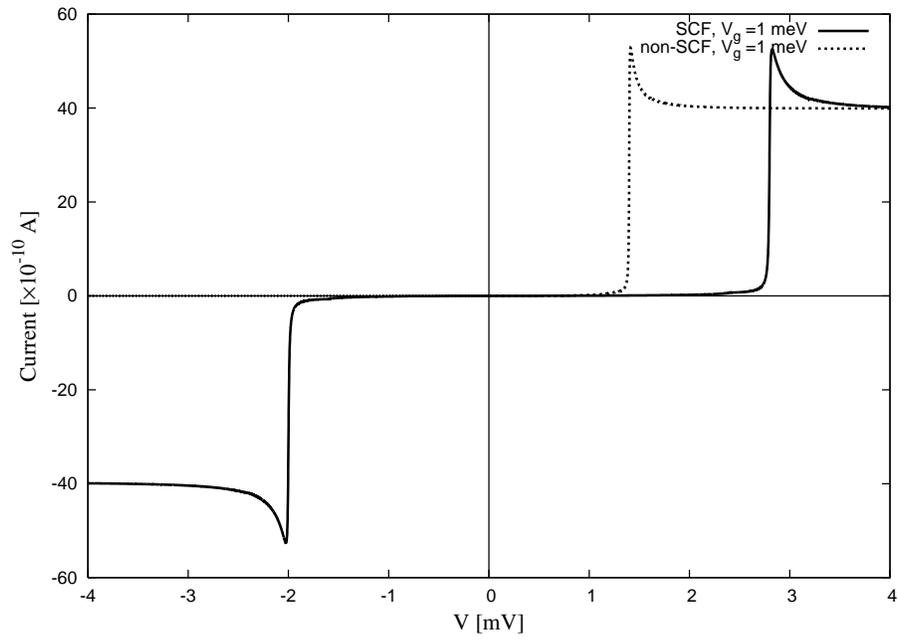}}
\caption{Zero temperature I-V characteristics calculated without the 
self  consistent field (non-SCF) method (dashed curve) and calculated 
using the self consistent field (SCF) method (solid curve), with 
$E_{d} = 0.2~$meV, $V_{g} = 1~$meV, $U_{0}=0.0025~$meV,
$C_{D}/C_{E} = 0.5$,
${\rm {\rm \Gamma}}_{L} = {\rm {\rm \Gamma}}_{R} = 0.008~$meV and
$\Delta = 0.2~$meV.}
\label{figure1}
\end{figure}

\begin{figure}
\resizebox{1.0\hsize}{!}{\includegraphics{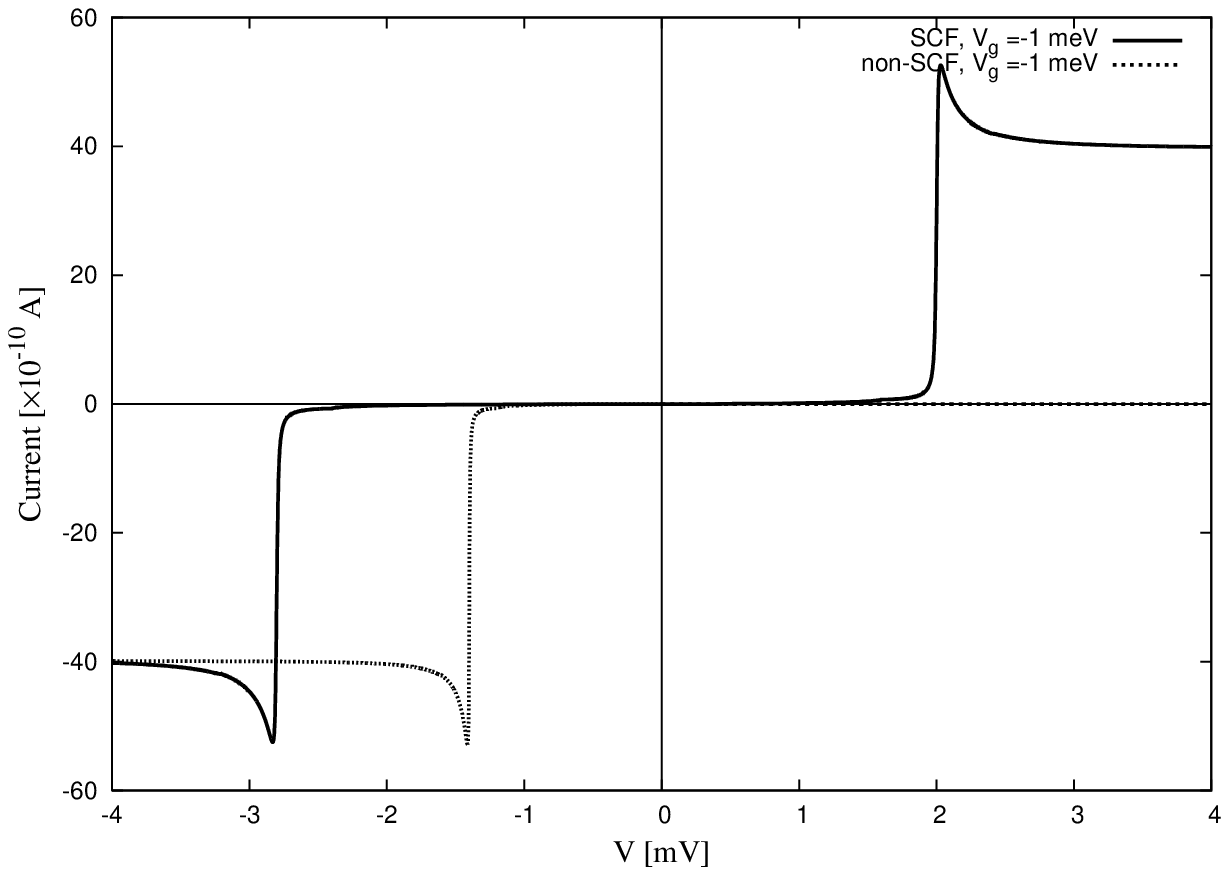}}
\caption{Zero temperature I-V characteristics calculated without the 
self  consistent field (non-SCF) method (dashed curve) and  calculated 
using the self consistent field (SCF) method (solid curve), with 
$E_{d} = 0.2~$meV, $V_{g} = -1~$meV, $U_{0} = 0.0025~$meV,
$C_{D}/C_{E}=0.5$,
${\rm {\rm \Gamma}}_{L} = {\rm {\rm \Gamma}}_{R} = 0.008~$meV and
$\Delta=0.2~$meV.}
\label{figure2}
\end{figure}

In the dashed curve in Figure \ref{figure1} we show zero temperature
${\rm I}-V$ characteristics, calculated without the self consistent
field (non-SFC) method for values of gate voltage $V_{g} > 0$.  In the
same figure the solid curve shows zero temperature ${\rm I}-V$
characteristics, calculated with the self  consistent field (SCF)
method for values of gate voltage $V_{g} > 0$ . As we can see in the
dashed curve, the current is nonzero for positive values of the drain
voltage, while for negative values of the drain voltage the current
vanishes out. For the solid curve the current can have nonzero values.

In Figure \ref{figure2} we show zero temperature ${\rm I}-V$
characteristics for gate voltage values $V_{g} < 0$ which are calculated
without the self consistent (non-SCF) method (dashed curve) and with
the self consistent method (SCF) (solid curve). In the first case
(non-SCF) the current is zero for positive values of the drain voltage
and nonzero for negative values of the drain voltage, while n the second
case (SCF) we can observe a symmetric ${\rm I}-V$ characteristic.

On the other hand, when we use the self consistent field (SCF) method,
the single particle current reaches the maximum value for higher drain
voltage as compared to the non-SCF method. It means that the presence of
the potential profile $U$ inhibits the electron flow. Furthermore, it is
noted that the $I-V$ characteristics, as calculated with the self
consistent field (SCF) method, agree with experimental data reported in
the literature \cite{PhysRevLett.74.3241}. In adittion, it proves that
the single particle electric current can have nonzero values for
positive and negative values of the drain voltage.

\section{Conclusions}\label{conclusions}
In conclusion, we have studied the effect of a self consistent
potential profile on single particle electric current across a
mesoscopic system attached to superconductor leads. Such system
describes a spin degenerated single quantum where Coulomb blockade is
neglected, in the regime $\Delta \gg {\rm {\rm \Gamma}}_{L,R}$.
We derived an exact single particle electric current by means of the
many body Keldysh technique. Zero temperature ${\rm I}-V$
characteristics agree with the experimental results. Furthermore,
we showed there are symmetric ${\rm I}-V$ characteristic, within the
self consistent method which address an interplay among potential
profile, single particle electric current and electronic population at
the mesoscopic region.


\begin{thebibliography}{22}
\bibitem{Baselmans}J. J. A. Baselmans, A. F. Morpurgo, B.J. Wees and
                   T. M. Klapwijk, Nature {\bf 397}, 43 (1999).
\bibitem{Volkov}A. F. Volkov, Phys. Rev. Lett. {\bf 74}, 4730 (1995).
\bibitem{Wilhelm}F. K. Wilhelm, G. Sch\"{o}n, A. D. Zaikin,  Phys. Rev.
                 Lett. {\bf 81}, 1682 (1998).
\bibitem{Yip}S. K. Yip,  Phys. Rev. Lett. {\bf 58}, 5803(1998).
\bibitem{Giazotto}F. Giazotto, T. T. Heikkil\"{a}, F. Taddei, R. Fazio,
                  J. P. Pekola and F. Beltram,  Phys. Rev. Lett. {\bf 92},
                  137001 (2004).
\bibitem{Pala}M. G. Pala, M. Governale and J. K\"{o}ing, New J. Phys. {\bf 9},
              278 (2007).
\bibitem{Defranceschi}, S. De Franceschi, L. Kouwenhoven, C. Sch\"onenberger and
                W. Wensdorfer, Nature Nanotechnology {\bf 5}, 703 (2010).
\bibitem{Fazio}R. Fazio and R. Raimondi, Phys. Rev. Lett. {\bf 80}, 2913 (1999).
\bibitem{Fazio1}R. Fazio and R. Raimondi, Phys. Rev. Lett. {\bf 82}, 4950 (1999).
\bibitem{PhysRevB.57.11891}K.  Kang,  Phys. Rev. {\bf B57}, 11891 (1998).
\bibitem{Schwab}P. Schwab and R. Raimondi, Phys. Rev. {\bf B59}, 1637 (1999).
\bibitem{Clerk}A. A. Clerk, V. Ambegaokar, S. Hershfield, Phys. Rev.
               {\bf B61}, 3555 (2000).
\bibitem{Shapira}S. Shapira, E. H. Linfield, C. J. Lambert, R. Seviour,
                 A. F. Volkov, A. V. Zaitsev, Phys. Rev. Lett.  {\bf 84},
                  159 (2000).
\bibitem{PhysRevB.55.R6137}Y. A. Levy, J. C. Cuevas,
               A. L\'{o}pez-D\'{a}valos, A. Mart\'\i{}n-Rodero, Phys. Rev. {\bf B55}, R6137 (1997).
\bibitem{Cuevas}J. C. Cuevas, Y. A. Levy, A. L\'{o}pez-D\'{a}valos,
                A. Mart\'\i{}n-Rodero, Phys. Rev.  {\bf B63}, 094515 (2001).
\bibitem{Liang}G. C. Liang, A. W. Ghosh, M. Paulsson, S. Datta,  Phys.
               Rev. {\bf B69}, 115302 (2004).
\bibitem{datta95}S. Datta, {\it Electronic Transport in Mesoscopic
                 Systems} (Cambridge University Press, UK, 1997).
\bibitem{keldysh65}L. Keldysh, Sov. Phys. JETP {\bf 20}, 1018 (1965).
\bibitem{haug}H. Haug and A. P. Jauho, {\it Quantum Kinetics in Transport
               and Optics of Semiconductors}  (Springer, Berlin, 1996).
\bibitem{PhysRev.124.41} P. W. Anderson, Phys. Rev. {\bf 124}, 41 (1961).
\bibitem{PhysRev.108.1175}J. Bardeen, L. N. Cooper and J. R. Schrieffer,
                          Phys. Rev. {\bf 108}, 1175 (1957).
\bibitem{langreth}D. C. Langreth, Phys. Rev. {\bf 150}, 516 (1966).
\bibitem{PhysRevLett.74.3241}D. C. Ralph, C. T. Black and M. Tinkham,
                             Phys. Rev. Lett. {\bf 74}, 3241 (1995).
\end{thebibliography}
\end{document}